\documentstyle[preprint,revtex,eqsecnum]{aps}
\begin{document}
\draft
\preprint{UAHEP-935}
\begin{title}
Non-Local Effects in String Black Holes
\end{title}
\author{B.Harms and Y.Leblanc}
\begin{instit}
Department of Physics and Astronomy, The University of Alabama\\
Box 870324, Tuscaloosa, AL 35487-0324
\end{instit}
\begin{abstract}
We consider modifications to general relativity due to non-
local string effects by using perturbation theory about the 4-
dimensional Schwarzschild black hole metric.  In keeping
with our interpretation in previous works of black holes as
quantum p-branes we investigate non-local effects due to a
critical bosonic string (1-brane) compactified down to 4
dimensions.  We show that non-local effects do not alter the
spacetime topology (at least perturbatively), but they do
lead to violations of the area law of black hole
thermodynamics and to Hawking's first law of black hole
thermodynamics.  We also consider a simple analytic
continuation of our perturbative result into the non-
perturbative region, which yields an ultraviolet-finite
theory of quantum gravity.  The Hawking temperature goes to
zero in the non-perturbative region (zero string tension
parameter), which is consistent with the view that
Planck-size physics is quantum mechanical.
\end{abstract}

\narrowtext
\section{Introduction}

In previous works ~\cite {BL92,BL93,HL93a,HL93b}, we
have shown that the semiclassical (WKB) approximation
of the Euclidean path integral of a quantized theory
of gravity in a black hole background should properly be interpreted
as the probability of quantum tunneling of a particle across the black
hole horizon barrier. This interpretation yielded directly the quantum
black hole degeneracy of states and resolved all the problems associated
with the traditional thermodynamical interpretation in a way consistent
with the laws of thermodynamics and quantum mechanics. We were then led
to the "finger-printing" of D-dimensional neutral black holes through
their degeneracy of states and to their identification as quantum
$(D-2)\over(D-4)$-branes.

Although our work has shed some light on the quantum black hole problem,
still questions remain unanswered. Of particular urgency is the classic
problem of field quantization in the classical black hole background.
It has been demonstrated over and over again that ``naive
quantization'' of fields in such
a background leads to an {\it exact} thermal
distribution for the particle number density with the (Hawking)
temperature related to the mass of the black hole, in agreement with
the WKB approximation calculation.

It has often been argued, and extensively investigated in various
models and dimensions, that quantum gravity effects (loop corrections
or back reaction effects) would actually resolve the problem of the
thermal spectrum. That it is indeed a problem is related to the
prediction that a black hole having absorbed initially particles in
pure states will emit them at later times in a thermal (mixed) state
during its evaporation process, thereby violating the law of quantum
mechanics asserting that pure states evolve solely into pure states
in any time-dependent process.

In this paper, we intend to investigate the nature of the above problem
through a series of (qualitative) arguments, partially supported
by explicit computations.

It is important to get a headstart on this problem and its resolution
by first accepting a present-day fact about quantum gravity, namely,
the only known "consistent" theories of quantized gravity are those
of quantum p-brane theories. In this work, as a simple working model,
we shall investigate critical bosonic string (1-brane) theory
compactified down to 4 dimensions. With the acceptance of such theories
as quantum gravity theories, comes the realization that general
relativity (GR) now becomes doubly modified, once through higher
derivative (non-local) classical contributions arising fundamentally
from the intrinsic non-local nature of p-branes, and once from purely
quantum (loop) corrections, the so-called back-reaction effects. In this
work we investigate the latter qualitatively and the former
quantitatively.

Non-local effects, such as (Riemann) curvature-square corrections
to Einstein's equations, have been considered by previous authors.
We nevertheless venture into such calculations in the following
section by applying perturbation theory about the 4-dimensional
Schwarzschild black hole metric. The small perturbation parameter
is taken to be the Regge slope $\alpha'\,=\,{1\over{2\pi{T}}}$,
where $T$ is the string tension. In agreement with previous results,
we find that the non-local $\alpha'$-corrections modify the so-called
area law $S\,=\,{A\over4}$ where $S$ is the Hawking entropy (Euclidean
action) and $A$ is the area of the black hole horizon. However,
contrary to previously published statements, we further find that
the so-called first law of Hawking's black hole thermodynamics is
also invalidated by such non-local classical effects. The
Hawking temperature is shown to be reduced by the $\alpha'$-corrections
and, pending plausible analytical continuations to strong $\alpha'$,
is shown to vanish in the limit $\alpha'\,\rightarrow\,\infty$. These
same analytical continuations are shown to also eliminate the black
hole singularity at the origin. The horizon (topological) structure
of the black hole spacetime remains however undisturbed by these
non-local effects, at least perturbatively. As will be discussed in a
later section, non-local effects remove the ultraviolet
divergences of quantum gravity in all Feynman loop diagrams involving
graviton propagators. This is a purely stringy effect and is directly
related to the removal of the black hole singularity
($r\,\rightarrow\,0$). With such corrections, gravity is shown to
vanish at the black hole center (asymptotic freedom). Finally, although
the Hawking temperature is modified by the $\alpha'$-corrections,
Green's functions of quantized fields in the modified black hole
background remain periodic with respect to the (modified) inverse
Hawking temperature (Kubo-Martin-Schwinger property) and so again
lead to an exact thermal spectrum for the particle number density,
even though the black hole singularity is removed.

Perturbatively, the classical stringy corrections to Einstein's
equations do not disturb the global topology of the spacetime.
This illustrates the point that the seemingly thermal property of
black holes is not intrinsic to the black hole as a curvature-
singular object. In fact, any regular spacetime, whether curved or
not and which possesses an horizon, leads to an exact thermal
spectrum (e.g. static de Sitter space, Rindler space, etc.).

Although non-local effects do not modify the thermal spectrum of
particles quantized in the classical black hole background, they
{\it do} invalidate, as mentioned earlier, certain aspects of the
traditional thermodynamical interpretation.

Clearly the resolution of the problem of the exact thermal spectrum
{\it must} lie within the quantum effects. Fundamentally, the origin
of the Hawking's radiation (thermal spectrum) phenomenon
lies in the fact that the horizon acts
as an impenetrable barrier between classically
causally disconnected sectors of spacetime and that, although a quantized
field may exist everywhere in the spacetime, information on the field
from sectors not attainable by an observer is seemingly lost.

Finding the true vacuum is a highly non-trivial exercise and is truly
a self-consistent (nonperturbative) problem in which the classical
background determines the quantization process which itself determines
this same background.

\narrowtext
\section{String Corrections to Black Hole Backgrounds}

In this section we calculate corrections to the metric
tensor due to `stringy' (non-local) classical effects.
String loop effects are ignored in this approximation. Including the
first order terms in the Regge slope expansion
gives for the corrected Einstein field equations ~\cite{green},
\begin{eqnarray}
R_{\mu\nu} + {\alpha'\over{2}}\;
R_{\mu\kappa\lambda\tau}R_{\nu}^{\kappa\lambda\tau} = 0\; .
\end{eqnarray}
This relationship is usually derived in higher dimensions (typically
$D=26$ for the critical bosonic string theory). We shall assume here
that all but four spacetime dimensions have been compactified with
the volume of the internal space taken equal to 1, for convenience.
We expand the metric tensor perturbatively in small
$\alpha'$ as follows,
\begin{eqnarray}
g_{\mu\nu} = g_{\mu\nu}^{(0)} + \alpha'\;
g_{\mu\nu}^{(1)}\;,
\end{eqnarray}
where $g_{\mu\nu}^{(0)}$ is the unperturbed
metric tensor and $g_{\mu\nu}^{(1)}$ is an unknown
function which we will obtain from substitution into
Eq.(2.1).
The metric tensor $g_{\mu\nu}^{(0)}$ is chosen to be of the
Schwarzschild form
\begin{eqnarray}
ds^2 = -(1 - {2\; M\over{r}})\;dt^2 + (1 - {2\;
M\over{r}})^{-1}\;dr^2 + r^2\; d\theta^2 + r^2\sin^2\theta\;
d\phi^2 \; .
\end{eqnarray}
We shall assume that the full metric tensor again describes a spherically
symmetric, static spacetime such that,
\begin{eqnarray}
g_{11} = -(g_{00})^{-1}\; .
\end{eqnarray}
To first order in $\alpha'$ Eq.(2.4) implies that,
\begin{eqnarray}
g_{11}^{(1)} = {g_{00}^{(1)}\over{(g_{00}^{(0)})^2}}
\end{eqnarray}
We also look for a solution such that $g_{22}^{(1)} = g_{33}^{(1)} =
0$.  Using the fact that
\begin{eqnarray}
R_{\mu\nu}^{(0)} = 0
\end{eqnarray}
and keeping only the lowest order terms in $\alpha'$, we
obtain the equation,
\begin{eqnarray}
R_{\mu\nu}^{(1)} - {1\over{2}}\;
R_{\mu\kappa\lambda\tau}^{(0)}\; R_{\mu}^{(0)\;
\kappa\lambda\tau} = 0\; .
\end{eqnarray}
As a result of our initial constraints on the solution we seek,
there is only one unknown function to be determined, namely
($g_{00}^{(1)}$). The set of the four equations (2.7) can be shown
to be self-consistent with the above requirements. These equations
give directly the perturbation of the Ricci tensor  from the knowledge
of the Riemann curvature of the unperturbed Schwarzschild black hole.
This in turn determines the perturbation of the metric.
Let
\begin{eqnarray}
g_{00}^{(1)} = a(r) f(r) \; ,
\end{eqnarray}
where,
\begin{eqnarray}
a(r) = 1 - {2\; M\over{r}}\; .
\end{eqnarray}
We find the following differential
equation for $f(r)$,
\begin{eqnarray}
{df\over{dr}} + {1\over{a(r)\;r}}f = -{6M^2\over{a(r)r^5}}\; .
\end{eqnarray}
The solution of this equation is easily found to be,
\begin{eqnarray}
f(r) = {2M^2\over{a(r)\;r}}\bigl( {1\over{r^3}} -
{1\over{r^3_+}}\bigr) \; ,
\end{eqnarray}
where ${r_+} = 2M$ is the position of the horizon of the unperturbed
black hole. The full metric tensor element $g_{00}$ thus has the
form, to order $\alpha'$,
\begin{eqnarray}
g_{00} = 1 - {{2M}\over{r}} - {{2M^2\alpha'\over{r}}
({1\over{r_+^3}} - {1\over{r^3}})}\; .
\end{eqnarray}
Setting $g_{00}$ to zero and solving for $r$, we find $r = {r_+}$,
which shows that the position of the
horizon is unchanged by the non-local effects.
The non-local $\alpha'$-effects therefore {\it do not} alter the
spacetime topology, at least from the viewpoint of perturbation
theory.

\section{String Corrections to the Thermodynamics of Black
Holes}

Having obtained non-local contributions to the metric tensor, we
can now investigate their effect on the thermally interpreted character
of the geometry.  According to Gibbons and Hawking ~\cite{gibb},
the thermodynamical properties of a black hole with inverse
temperature $\beta_H$ can be obtained by analytically
continuing the metric to Euclidean spacetime, and summing
over geometries which are asymptotically flat, have topology
$R^2 \times S^2$ and are periodic in imaginary time $\tau$
with period $\beta_H$.  The thermodynamical quantities of
interest are obtained from the partition function, which in
the usual semiclassical (WKB) approximation is given by,
\begin{eqnarray}
Z(\beta_H) \sim \exp (-S_E/\hbar) \; ,
\end{eqnarray}
where $S_E$ is the Euclidean action evaluated at the black hole
solution.  The total Euclidean action
is the sum of the analytically continued Einstein-Hilbert action,
\begin{eqnarray}
S = -{1\over{16\pi}}\int d^4x \sqrt{g} R + S_{boundary}\; ,
\end{eqnarray}
and the contribution to the action from the nonlocal effects (to
first order in $\alpha'$),
\begin{eqnarray}
S_{nonlocal} = -{\alpha'\over{64\pi}}\int d^4x \sqrt{g}
{R_{\mu\nu\sigma\lambda}}{R^{\mu\nu\sigma\lambda}}\; .
\end{eqnarray}
Since the total action $S_E$ in Eq.(3.1) is to be evaluated at
the black hole solution of Einstein's equations (Eq.(2.1)),
we find,
\begin{eqnarray}
S_E \simeq -{1\over{32\pi}}\int d^4x \sqrt{g} R + {S_{boundary}}\;,
\end{eqnarray}
which is valid up to first order in $\alpha'$-perturbation theory.

With regards to the curvature term of the above equation, the results
from the previous section show that the only non-vanishing contributions
are those originating from the non-local effects. One finds
$S_o \simeq \alpha'{{\beta_H}\over{16{M^2}}}$, where $S_o$ is the
action of the curvature term in Eq.(3.4).

Now the boundary term in Eq.(3.4) is evaluated as follows,
\begin{eqnarray}
S_{boundary} = -{1\over{8\pi}}\partial_{normal}(volume\;
of\; boundary) \; ,
\end{eqnarray}
For a Euclidean spacetime with a
metric of the form
\begin{eqnarray}
ds^2 = \lambda^2(r)\;d\tau^2 + \lambda^{-2}(r)\; dr^2 + r^2\;
d\Omega^2 \; ,
\end{eqnarray}
the volume of the boundary with $S^1 \times S^2$ topology
is given as follows,
\begin{eqnarray}
V = [\lambda(r) \beta_H] 4\pi r^2\; .
\end{eqnarray}
The boundary action evaluated from Eq.(3.5) is now given by,
\begin{eqnarray}
S_{boundary} = -{\beta_H\over{2}}
\lambda(r)\;
{d\over{dr}}(r^2\lambda(r))_{r\rightarrow\infty}\; .
\end{eqnarray}
As in the case of $\alpha' = 0$, $S_{boundary}$ diverges as
$r \to \infty$.  This problem is remedied by Gibbons and
Hawking ~\cite{gibb} by subtracting the corresponding action of a flat
spacetime at the boundary.  The
term to be subtracted has the following form,
\begin{eqnarray}
S_{boundary}^{flat} = -{\beta_H\over{2}}\lambda(r)
{d\over{dr}}(r^2)_{r\rightarrow\infty}\; .
\end{eqnarray}
Explicit calculations yield the following result,
\begin{eqnarray}
S_{boundary} - S_{boundary}^{flat} = {\beta_H}
\bigl({M\over2} + {\alpha'\over{16{M^2}}}\bigr)\;.
\end{eqnarray}
Adding the regulated boundary term and the non-local contributions
$S_o$ gives a total action
\begin{eqnarray}
S_E = \beta_H ({M\over{2}} +
{1\over{8}}{\alpha'\over{M^2}}) \; .
\end{eqnarray}
The requirement of the vanishing of the conical singularity
of the Euclidean spacetime yields the inverse Hawking temperature,
\begin{eqnarray}
\beta_H = {2\pi}\bigl[
{d\over{dr}}(\lambda(r))\;\lambda(r)\bigr]^{-1}_{r={r_+}}\; .
\end{eqnarray}
Explicitly, we find,
\begin{eqnarray}
\beta_H = {8\pi M}\bigl(1 + {{3\alpha'}\over{8{M^2}}}\bigr)\;,
\end{eqnarray}
which to order $\alpha'$ yields a total action (Hawking entropy) of,
\begin{eqnarray}
S_E = 4\pi M^2(1 + {5\alpha'\over{8M^2}}) \; .
\end{eqnarray}
In terms of the area of the horizon,
\begin{eqnarray}
S_E = {A\over{4}}(1 + {5\alpha'\over{8M^2}}) \; ,
\end{eqnarray}
where $A = 4\pi{r_+^2}$ is the area of the horizon.

This result shows that if non-local effects are included in
Einstein's equations, the area law of black hole
thermodynamics ~\cite{bek,hawk1} is violated. Similarly the canonical
temperature is not equal to the Hawking temperature,
\begin{eqnarray}
\beta = {dS_E\over{dM}} = 8\pi M \; ,
\end{eqnarray}
a result to be compared with Eq.(3.13) above. This is a violation
of Hawking's first law of black hole thermodynamics ~\cite{hawk2}.
This result disagrees with some previously
published calculations in the context
of superstring theory ~\cite{Meyer}.

\narrowtext
\section{Ultraviolet Finite Quantum Gravity}

The content of this section is primarily based upon the Limiting
Curvature Hypothesis (LCH) ~\cite{Brand}
which states that no classical curvature
singularity should occur in a suitable geometrical gravitational
theory. This is required in order for the spacetime to be geodesically
complete. This principle effectively provides for the realization
of Penrose's Cosmic Censorship Hypothesis.

Another important input into the working assumptions of this section
is the belief that the non-local p-brane theories (including of
course string theories) are ultraviolet (UV) finite theories.

As is clear from the viewpoint of our solution Eq.(2.12) for the
black hole metric incorporating the non-local effects to first
order in $\alpha'$, the singular behavior at the origin of the
metric worsens as we go to higher orders in $\alpha'$-perturbation
expansion. That it is so originates from the fact that the higher order
terms in the corrected Einstein's equations have an increasingly
higher number of derivatives. The perturbation series in $\alpha'$
is therefore a series of individually divergent terms as $r \rightarrow
0$. However, in view of the arguments of the preceding paragraphs,
the series itself should be regular as $r \rightarrow 0$. To uncover
the full solution would probably require a great deal more calculation
involving the higher order terms.

For our purposes however, a simple analytic continuation of the
result of Eq.(2.12)  to strong $\alpha'$ (although it is by no means
unique) will suffice. Let us consider the following analytical
continuation of Eq.(2.12) to strong $\alpha'$ (or small $r$),
\begin{eqnarray}
g_{oo}\;=\;1\;-\;{{2M}\over{r}}\,e^{{\alpha'}M({1\over{r_+^3}} - {1
\over{r^3}})}\;.
\end{eqnarray}

Notice that the horizon of this spacetime is again left undisturbed,
namely $r_+\,=\,2M$.

It is now apparent that the singularity at the origin has been removed.
Indeed spacetime is flat at the origin. Now since $g_{oo}\,=\,1 + 2U(r)$,
in which $U(r)$ is the gravitational potential, Eq.(4.1) gives the
following stringy modification to Newton's gravity,
\begin{eqnarray}
U(r)\;=\;- {M\over{r}} e^{{\alpha'}M({1\over{r_+^3}} - {1\over{r^3}})}\;.
\end{eqnarray}
It may be useful to comment here that the non-local $\alpha'$-corrections
to Newton's law will produce modifications (although very small) to
phenomena such as the gravitational redshift, the bending of light by
the sun and the perihelion of Mercury. If refined experiments were
carried out (which may be decades or centuries away), these would allow
for the first direct measurements of the string tension.

Let us now express the Fourier transform of
the gravitational potential $U(r)$ as follows,
\begin{eqnarray}
U(k)\;\equiv\;- {M\over{4\pi}}
{\int_{-\infty}^{\infty}} d^3{\vec x}
\bigl\{
{ {|\vec x|}^{-1} \exp{\bigl[{\alpha'}
M \bigl({1\over{8{M^3}}} - {1\over{|\vec x|^3}}
\bigr)\bigr]\bigr\} e^{i{\vec k}\cdot{\vec x}}\;\;;
(k = |\vec k|)\;.
\end{eqnarray}
We find,
\begin{eqnarray}
U(k)\;=\;- \lim_{\epsilon \rightarrow {0}} \int_0^{\infty} dr
r^{1-\epsilon} {\sin{kr}\over{kr}} \exp{\bigl[{\alpha'}M\bigl(
{1\over{8M^3}} - {1\over {r^3}}\bigr)\bigr]\;,
\end{eqnarray}
in which $\epsilon$ is an infrared (IR) ($r\rightarrow\infty$) regulator.

Introducing the dimensionless parameter $\sigma \equiv kr$, we finally
arrive at the following explicit integral representation for the
graviton ``propagator'' $U(k)$,
\begin{eqnarray}
U(k)\;=\;- {1\over{k^2}} e^{{\alpha'}\over{8M^2}} \lim_{\epsilon
\rightarrow 0} \int_0^{\infty} d\sigma\,{{\sin{\sigma}}\over{\sigma^
{\epsilon}}}\,e^{{-{\alpha'}Mk^3}\over{\sigma^3}}\;.
\end{eqnarray}

Two important remarks must be made. First, as is clear from the
above formula, the IR behavior of the graviton propagation function
remains unaltered by the (re-summed) non-local $\alpha'$-effects.
The graviton remains a massless particle and no ghost particles occur
in this expression. The second remark of course concerns the UV
behavior of the propagator, which clearly vanishes exponentially
as $k \rightarrow \infty$. It is now obvious that any loop Feynman
diagrams involving internal graviton propagators lead to UV-finite
results. Loop diagrams should be evaluated by first integrating
over the loop momenta, then integrating over the integral representation
parameters $\sigma_i$'s and finally taking the $\epsilon_i$'s IR
regulators to zero, in that order.

Of course these apply to the Feynman rules of a field theory of gravity,
namely quantum GR, which itself is the ${\alpha'}\rightarrow 0$ limit
of string theory. The above considerations show that not all
${\alpha'}$-effects should be taken to zero in order to recover the
field theory limit, if finite predictions are to be made from such
a theory.

Finally we close this section by mentioning a possible continuation of
Eq.(3.13) for the Hawking temperature to the strong $\alpha'$-domain
(or equivalently to Planck mass black holes)

Let us generalize Eq.(3.13) as follows,
\begin{eqnarray}
\beta_H \;=\; 8{\pi}{M}\,e^{{3{\alpha'}}\over{8M^2}}\;.
\end{eqnarray}
Clearly (pending the plausibility of this argument), the Hawking
temperature goes to zero as ${\alpha'}\rightarrow \infty$. This behavior
seems consistent with the view that Planck size physics should belong
to the realm of quantum mechanics, if one
starts from Hawking's thermodynamical viewpoint on black holes. Of course,
if one belongs to the so-called Quantum School, such an argument
is really unnecessary, as black holes of all sizes and masses are
viewed as pure quantum excitations of some quantum gravity theory.

\narrowtext
\section{Discussion}

In this work, we addressed the problem of the effects of the
classical non-local ($\alpha'$-corrections) modifications of GR from
the string progenitor, in connection with black hole
spacetimes.

Our results are that the non-local effects do not disturb the topological
character of the spacetime  (the horizon remains unshifted, at least
perturbatively), while nevertheless invalidating the traditional
thermodynamical interpretation of the black hole spacetime.

In addition, pending the plausible analytical continuations of the
results of Section II to the strong $\alpha'$ domain, as discussed in
Section IV, we found that the quantum field theory limit (quantum GR)
of string theory is UV-finite, provided some $\alpha'$-dependency
remains in the field theory limit.

On the other hand, because the non-local classical $\alpha'$-corrections
do not influence the topology of the black hole (although they do
reduce its Hawking temperature), field quantization in the black
hole background will continue to yield a thermal spectrum (with
shifted Hawking temperature) for the particle number density, a
somewhat disappointing result if one is to believe in the pure quantum
nature of black holes.

As mentioned in the Introduction, the resolution of the
thermal spectrum must originate from the quantum or back reaction
effects. Attempts to resolve this problem have been made by various
authors and in various numbers of dimensions.

An intuitive, if not totally naive, way to see that this may indeed
occur, is the following consideration making use of the statistical
bootstrap model of quantum Schwarzschild black holes. The situation
is depicted in Fig.~\ref{boot}, where a single massive quantum black hole
is modeled as a gas consisting itself of a single quantum massive
black hole excitation surrounded by countless extreme (massless)
others. Of course the equilibrium state of this gas is not thermal,
as the energy distribution is highly inhomogeneous. If for some
reason the massive excitation is cut-off from the rest, the original
quantum black hole will resemble a gas of massless excitations
in thermal equilibrium. It is the effect of the single massive
excitation in the gas which drastically changes the nature of the
equilibrium configuration and allows
for the bootstrap property. This massive
excitation of course is a pure quantum gravity (back reaction)
effect of the p-brane type. Should such effects be included in the
usual semiclassical treatments, the above statistical model
shows that the thermal spectrum problem would be resolved
satisfactorily by yielding a picture of the original black hole
as a pure state. We intend to address this problem much more deeply
in future publications.

As a final comment, it is very sensible to expect quantum gravity
effects to induce an effective cosmological constant in the original
metric as loop corrections induce a non-zero vacuum energy.
Consequently, it may very well be that there is no such a thing as
a quantum Schwarzschild black hole. The metric instead would be of the
Schwarzschild-de Sitter type, which is known to contain at most two
real horizons. This quantum-induced change of the topology of the
black hole spacetime may have drastic effects on the thermal
distribution of the particle number density. Explicit computations
are currently being carried out.

\acknowledgments

This research was supported in part by the U.S. Department
of Energy under Grant No. DE-FG05-84ER40141.

\figure{Statistical Bootstrap : $\Omega(M)\,\sim\,\rho(M)\;\;(M
\rightarrow\infty)\,.\label{boot}}
}
\end{document}